\documentclass{IEEEtran4PSCC}

\ifCLASSINFOpdf
\else
\fi

\usepackage[cmex10]{amsmath}

\usepackage[T1]{fontenc}
\usepackage[utf8]{inputenc}
\usepackage[hyperindex=true, %
  pdftitle={Experimental Validation of Feedback Optimization in Power Distribution Grids},%
  pdfauthor={Lukas Ortmann, Adrian Hauswirth, Ivo Caduff, Florian Dörfler,  Saverio Bolognani},%
  colorlinks=true,%
  pagebackref=false,%
  plainpages=false,%
  pdfpagelabels,%
  linkcolor=black,%
  citecolor=black,%
  filecolor=black,%
  urlcolor=black%
]{hyperref} % colourlinks=false for printing

\makeatletter
\let\old@ps@headings\ps@headings
\let\old@ps@IEEEtitlepagestyle\ps@IEEEtitlepagestyle
\def\psccfooter#1{%
    \def\ps@headings{%
        \old@ps@headings%
        \def\@oddfoot{\strut\hfill#1\hfill\strut}%
        \def\@evenfoot{\strut\hfill#1\hfill\strut}%
    }%
    \def\ps@IEEEtitlepagestyle{%
        \old@ps@IEEEtitlepagestyle%
        \def\@oddfoot{\strut\hfill#1\hfill\strut}%
        \def\@evenfoot{\strut\hfill#1\hfill\strut}%
    }%
    \ps@headings%
}
\makeatother

\usepackage[dvipsnames]{xcolor}
\usepackage{tikz}
\usepackage{pgfplots} 
\usepackage{pgfgantt}
\usepackage{pdflscape}
\pgfplotsset{compat=newest} 
\pgfplotsset{plot coordinates/math parser=false}
\usepackage{wrapfig}
\usepackage{multirow}
\usepackage{caption}
\usepackage{amssymb}
\usepackage{cite}
\usepackage{algorithm}
\usepackage[noend]{algpseudocode}
\usepackage[absolute,overlay]{textpos}

\usepackage[font=small]{caption}

\newlength\fwidth
\newlength\fheight
\usepackage[prependcaption,colorinlistoftodos]{todonotes}

\usepackage{booktabs}

\begin{document}
\title{Experimental Validation of Feedback Optimization in Power Distribution Grids}

%% To specify the authors when (number of affiliations <= 2)
\author{
\IEEEauthorblockN{Lukas Ortmann, Adrian Hauswirth, Ivo Caduff, Florian D{\"o}rfler, Saverio Bolognani}
\IEEEauthorblockA{%Automatic Control Laboratory\\
ETH Zurich, 8092 Zurich, Switzerland}%\\
%\{ortmannl,hadrian,dorfler,bsaverio\}@ethz.ch
}

%% To specify the authors when (number of affiliations > 2)
% \author{\IEEEauthorblockN{Author n.1\IEEEauthorrefmark{1},
% Author n.2\IEEEauthorrefmark{2},
% Author n.3\IEEEauthorrefmark{3}, 
% Author n.4\IEEEauthorrefmark{3} and
% Author n.5\IEEEauthorrefmark{4}}
% \IEEEauthorblockA{\IEEEauthorrefmark{1} Department Name of Organization A\\
% Name of the organization A,
% Address A\\ Emails if wanted}
% \IEEEauthorblockA{\IEEEauthorrefmark{2} Department Name of Organization B\\
% Name of the organization B,
% Address B\\ Emails if wanted}
% \IEEEauthorblockA{\IEEEauthorrefmark{3} Department Name of Organization C\\
% Name of the organization C,
% Address C\\ Emails if wanted}
% \IEEEauthorblockA{\IEEEauthorrefmark{4}Department Name of Organization D\\
% Name of the organization D,
% Address D\\ Emails if wanted}
% }

\maketitle


\begin{textblock*}{\textwidth}(15mm,10mm) % {block width} (coords) 
	\centering \bf \textcolor{NavyBlue}{Published on \emph{Electric Power Systems Research,} vol. 189, December 2021.\\\url{https://doi.org/10.1016/j.epsr.2020.106782}}
\end{textblock*}

\begin{abstract}
We consider the problem of controlling the voltage of a distribution feeder using the reactive power capabilities of inverters. On a real distribution grid, we compare the local Volt/VAr droop control recommended in recent grid codes, a centralized dispatch based on optimal power flow (OPF) programming, and a feedback optimization (FO) controller that we propose.
The local droop control yields suboptimal regulation, as predicted analytically.
The OPF-based dispatch strategy requires an accurate grid model and measurement of all loads on the feeder in order to achieve proper voltage regulation.
However, in the experiment, the OPF-based strategy violates voltage constraints due to inevitable model mismatch and uncertainties.
Our proposed FO controller, on the other hand, satisfies the constraints and does not require load measurements or any grid state estimation. The only needed model knowledge is the sensitivity of the voltages with respect to reactive power, which can be obtained from data. As we show, an approximation of these sensitivities is also sufficient, which makes the approach essentially model-free, easy to tune, compatible with the current sensing and control infrastructure, and remarkably robust to measurement noise. 
We expect these properties to be fundamental features of FO for power systems and not specific to Volt/VAr regulation or to distribution grids.
\end{abstract}


% Use this to place sponsorships
\thanksto{
\setlength{\columnsep}{6pt}%
\setlength{\intextsep}{3pt}%
\begin{wrapfigure}{o}{0.14\columnwidth}
  \raggedleft
  \vspace{-4pt}
    \includegraphics[width=0.14\columnwidth]{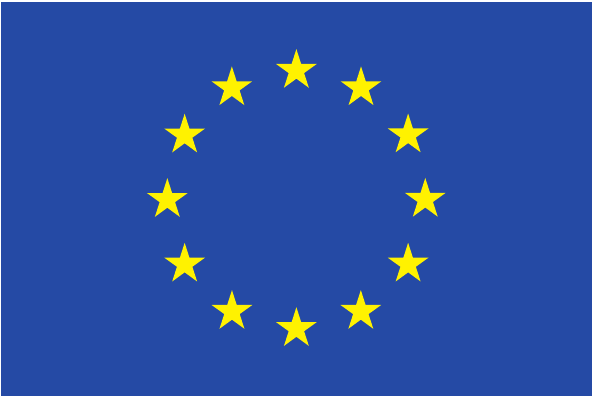}
\end{wrapfigure}
This research has been performed using the ERIGrid Research Infrastructure and is part of a project that has received funding from the European Union's Horizon 2020 Research and Innovation Program under the Grant Agreement No. 654113. The support of the European Research Infrastructure ERIGrid and its partner Technical University of Denmark is very much appreciated.
This paper reflects only the authors' view and the EU Commission is not responsible for any use that may be made of the information it contains.\\
\hspace*{\parindent} The research leading to this work was supported in part by the Swiss Federal Office of Energy grant \#SI/501708 UNICORN.\\
\hspace*{\parindent} Corresponding author: Lukas Ortmann, email: \href{mailto:ortmannl@ethz.ch}{ortmannl@ethz.ch}.
}

\section{Introduction}

The shift towards distributed microgeneration and the change in power consumption (electric mobility, storage, flexible loads) poses unprecedented challenges to power distribution grids. One important concern is the occurrence of over- and undervoltages in distribution feeders, which may force the distribution system operator to curtail generation or to shed loads, respectively. 
The flexibility of the power inverters of distributed energy resources (DERs), and more precisely their reactive power capabilities, can be used to avoid these extreme remedial actions. Control of reactive power flows is a relatively inexpensive way to regulate the feeder voltage and should therefore be fully exploited in order to avert taking action on the active power flows in the grid.

Many \emph{local control strategies} have been proposed towards this goal. In these strategies, each DER only measures the voltage at its point of connection in order to decide its own reactive power set-point. No communication infrastructure is needed because the controllers are fully decentralized. An example of local control strategies are static control laws like droop control with dead band and saturation, which have been included in the recent grid codes \cite{IEEE1547,VDE,ENTSOE}. 
Incremental local control strategies have also been proposed, where the reactive power set-point is calculated as a function of the voltage magnitude and the past reactive power set-point \cite{Cavraro2017,Farivar2015}. 
The main advantage of local control strategies is that they are easy to implement due to being fully decentralized. However, it was recently shown that they are suboptimal \cite{Bolognani2019}. Namely, they do not necessarily regulate the voltage to the admissible range, even with sufficient reactive power capability of the inverters.

An alternative solution to the voltage regulation problem is to use an optimal power flow (OPF) solver to calculate the optimal reactive power set-points (see \cite{EVANGELOPOULOS201695} and references therein). This optimization-based method requires an accurate grid model and full observability of the grid state, neither of which are usually available in distribution grids. Estimating the real-time state of a distribution grid is only possible if enough sensors are deployed which adds significant complexity and cost to this approach.

A third and more promising option is \emph{feedback optimization} (FO) or \emph{autonomous optimization}. FO has been recently proposed as a strategy to adjust DER set-points in real-time and to drive the system to an optimal operating point without measuring or estimating the power demands \cite{Bolognani2015,Dallanese2016,Hauswirth2017,Qu2018,Magnusson2019,Molzahn2017,Low2017,Dallanese2019}.
To the best of the authors' knowledge, there is no publicly available report of testing of these solutions on a real grid, and their robustness to model mismatch and measurement noise has been conjectured but never verified in experiments.

This paper presents an experimental verification of the effectiveness of FO for Volt/VAr control on a simple, yet plausible testbed. The experiment shows that the grid state converges to the optimal reactive power flow, and it allows to assess the performance in the presence of: 
\begin{itemize}
    \item \emph{model mismatch}, especially in comparison to standard OPF-based dispatch, showing that FO performs well with an extremely rudimentary model of the grid;
    \item realistic \emph{measurement accuracy}, based on off-the-shelf sensors and without any state-estimation stage.
\end{itemize}
Additionally, the experiment illustrates the suboptimality of the local Volt/VAr control strategies included in recent grid codes. As predicted in \cite{Bolognani2019}, they can be ineffective and even detrimental in regulating under- and overvoltages, leading to more loads being shed or renewable generation being curtailed than necessary.

The rest of the paper is structured as follows: In Section~\ref{sec:FO} the general concept of FO is presented and the assumptions are introduced that make the implementation more tractable. Afterwards a FO controller is designed for the Volt/VAr problem. The experimental setup and the controller implementation are explained in Section~\ref{sec:experimental_setup} and Section~\ref{sec:controller_implementation}, respectively. Finally, the experimental results are presented in Section~\ref{sec:experiments} and the paper is concluded in Section~\ref{sec:conclusion}.

\section{Feedback Optimization}
\label{sec:FO}

Consider the problem of determining the values of some set-points $u$ (e.g, reactive power injections) in order to minimize a given cost function (typically a cost of the control effort) while satisfying some constraint on an output signal $y$ (e.g, voltage bounds).
The output $y$ is also affected by an exogenous uncontrollable input $w$ (e.g, power demand of the loads), and depends on these inputs via a nonlinear map $y=h(u,w)$.

The aforementioned decision problem is mathematically represented by the possibly non-convex optimization problem
\begin{equation}
\begin{aligned}
		\min_u &\ f(u) 
		&& \text{cost of actuation effort}\\
		\text{s.t.}\; & \ g(y)\leq 0 
		&& \text{constraints on the output $y=h(u,w)$}\\
		&\ u \in \mathcal{U}
		&& \text{actuation bounds.}
\end{aligned}
\label{eq:generic_optimization_problem}
\end{equation}
For a more general approach with $f(u,y)$ and consideration of underlying dynamics, see \cite{Hauswirth2019}.
One way to approach this decision problem is to solve \eqref{eq:generic_optimization_problem} using the model $y=h(u,w)$ and then apply the resulting set-points to the system in a feedforward manner. This approach comes with several disadvantages, such as the need for an accurate model $h$ of the system and for full measurement or an estimate of the exogenous input $w$.

An alternative approach is called \emph{feedback optimization}, and is based on the assumption that the output $y$ of the system can be measured in real-time, while the exogenous input $w$ is unmonitored.
Real-time measurements are used to iteratively adjust the set-points $u$, based on reduced model information, in such a way that the closed-loop system converges to the solutions of the optimization problem \eqref{eq:generic_optimization_problem} (hence the name).

\subsection{Feedback Optimization Principle}
\label{subsec:FO_principle}

The core idea behind FO is to exploit the measurements $y$ instead of relying on the model $y=h(u,w)$.
One way to do so is to dualize the output constraints and get the Lagrangian
\begin{equation}
	\mathbf{L}(u,\lambda)=f(u)+\lambda^T g(h(u,w)) ,
\end{equation}
where $\lambda$ is a vector of dual variables in which each dual variable corresponds to one constraint.
Instead of \eqref{eq:generic_optimization_problem} we consider the  optimization problem
\begin{equation}
		\max_{\lambda \ge0} \ \phi(\lambda) ,
	\label{eq:dual_problem}
\end{equation}
where the dual function $\phi(\lambda)$ is defined as
\begin{equation}
        \phi(\lambda) := \min_{u\in \mathcal U}\; \mathbf{L}(u,\lambda) .
        \label{eq:subproblem}
\end{equation}
Assuming that the feasible space of \eqref{eq:generic_optimization_problem} has a non-empty interior, \eqref{eq:generic_optimization_problem} and \eqref{eq:dual_problem} have the same solution (Strong Duality Theorem, \cite[Proposition 5.3.1]{Bertsekas1999}). 

To solve \eqref{eq:dual_problem} we use a gradient ascent with a fixed step size, in which the multiplier $\lambda$ is repeatedly updated in the direction of steepest ascent of $\phi(\lambda)$, while ensuring $\lambda \ge 0$.
By introducing the element-wise projection operator 
$
[a]_{\geq 0}:=\max\{a,0\}
$
and the tuning parameter $\alpha$ we can write
\begin{equation}
	\lambda(t+1) = [\lambda(t)+\alpha \nabla_\lambda \phi(\lambda)]_{\geq 0}.
	\label{eq:dualascent}
\end{equation}
In \cite[Proposition 6.1.1]{Bertsekas1999} it was shown that $\nabla_\lambda \phi(\lambda) =g(h(u,w))$. In other words, the gradient of $\phi$ is given by the violation of the dualized constraints $g(h(u,w))$ at the solution of the optimization problem \eqref{eq:subproblem}, leading to:
\begin{equation}
	\lambda(t+1) = [\lambda(t)+\alpha g(y(t))]_{\geq 0}.
    \label{eq:generic_lambda_update}
\end{equation}
Instead of computing $g(y(t))$ based on model information, we exploit the physical system to enforce the constraint $y=h(u,w)$ and measure the output $y=h(u,w)$ as feedback from the plant. The variable~$\lambda$ integrates the output constraint violation with a step size of~$\alpha$. Note, that this corresponds to the integral part of a PI-controller.

Using $\lambda(t+1)$, we update the set-points $u$ with the solution of \eqref{eq:subproblem}, i.e., 
\begin{equation}
    \begin{split}
        u(t+1)&=\arg\min_{u \in\mathcal U} \; \mathbf{L}(u,\lambda(t+1))\\
    &=\arg\min_{u \in\mathcal U} \; f(u)+\lambda(t+1)^T g(h(u,w)).
    \end{split}
\label{eq:explicitoptimization}
\end{equation}
Whether this optimization problem is easier to solve than the original one in \eqref{eq:generic_optimization_problem} is not apparent at this point.
In the next subsection we will see how, under mild assumptions, this optimization problem admits an approximation which is numerically very tractable.

To summarize, the FO controller is realized by running the following algorithm at every time $t=0,1,\ldots$
\begin{algorithm}
\caption{Feedback optimization controller}
\begin{algorithmic}[1]
    \State Measure the system output $y(t)$
    \State Calculate $\lambda(t+1)$ as in \eqref{eq:generic_lambda_update}
    \State Solve the optimization problem in \eqref{eq:explicitoptimization}
    \State Apply the calculated set-points $u(t+1)$ to the system
\end{algorithmic}
\end{algorithm}

See Figure~\ref{fig:block_diagram} for a block diagram of a FO controller for the Volt/VAr problem, that we derive in Section \ref{subsec:Volt/VAr}.

\subsection{Practical Feedback Optimization Design}
\label{subsec:practical_FO}
We now make two assumptions that are not necessary, but make the FO controller numerically more tractable. First, we assume the cost is a quadratic function $f(u)=\tfrac{1}{2}u^TMu$ with $M$ being square, symmetric and positive semidefinite.
Second, we make the mild assumption that the constraints on the input and output are linear. We therefore get $\mathcal{U}=\{u \; | \; Cu\leq d \}$ and $g(y)=Ay-b$. Linearity of the constraints is often given, as in many cases the limits are upper and lower bounds of the form $u_\text{min} \le u \le u_\text{max}$.
This leads to \eqref{eq:generic_optimization_problem} taking the form
\begin{equation}
    \begin{aligned}
        \min_{u} &\ \dfrac{1}{2}u^TMu && \text{quadratic cost of actuation}\\
        \text{s.t.} &\  A y \leq b && \text{linear constraints on the output $y=h(u,w)$} \\
                    &\ C u \le d && \text{linear actuation bounds.}
    \end{aligned}
    \label{eq:optimization_problem_qp}
\end{equation}
Notice that the output is still a possibly nonlinear and non-convex function of the input $y=h(u)$.
The dual update \eqref{eq:generic_lambda_update} for the special case \eqref{eq:optimization_problem_qp} of \eqref{eq:generic_optimization_problem} takes the form:
\begin{equation}
	\lambda(t+1) = [\lambda(t)+\alpha (Ay(t) - b)]_{\geq 0}.
    \label{eq:lambda_update_linear}
\end{equation}
However, the major advantage of \eqref{eq:optimization_problem_qp} over \eqref{eq:generic_optimization_problem}, lies in the evaluation of \eqref{eq:explicitoptimization} which can now be explicitly solved.
There are several ways to solve \eqref{eq:explicitoptimization}. We choose to do this in two steps that we feel are easy to understand. First, we ignore the constraint $u \in \mathcal{U}$ and calculate the critical point $u$ for which $\nabla_u \mathbf{L}(u,\lambda(t+1))=0$ (first order optimality condition). Then, we project this unconstrained critical point onto $\mathcal{U}$. 

The derivative of the Lagrangian $\mathbf{L}(u,\lambda(t+1))$ is
\begin{equation}
\begin{split}
    	\nabla_u \mathbf{L}(u,\lambda(t+1))
		&\!=\! \nabla_u f(u)\!+\! \nabla_u \left( \lambda^T(t+1) g(h(u))\right)\\
		&= Mu + {\dfrac{\partial h(u,w)}{\partial u}}^T A^T \lambda(t+1).
\end{split}
\end{equation}
The factor $\tfrac{\partial h(u,w)}{\partial u}$ is the sensitivity of the output $y$ with respect to the input $u$. This sensitivity is in general dependent on $u$ and $w$, but in many practical applications can be approximated by a constant matrix $H$. Furthermore, the approximation error will be compensated by the feedback nature of this scheme.
The theoretical analysis of this robustness remains an open question, and is one of the main motivations for the experimental validation reported in this paper.
Under this modeling assumption we have
\begin{equation}
	\nabla_u \mathbf{L}(u,\lambda(t+1)) \approx Mu + H^T A^T \lambda(t+1),
\end{equation}
and a critical point of $\mathbf{L}(u,\lambda(t+1))$ in $u$ is approximated by
\begin{equation}
u_\textrm{unc} := -M^{-1} H^T A^T \lambda(t+1).
\label{eq:u_unc}
\end{equation}
This is the unconstrained critical point.
The solution to the constrained case is obtained by projecting $u_\textrm{unc}$ onto the set of feasible control inputs $\mathcal{U}$, that is
\begin{equation}
\begin{split}
    u(t+1) &= \arg \min_{u \in \mathcal{U}} \  \|u-u_\textrm{unc}\|_M^2\\
     &= \arg \min_{u \in \mathcal{U}} \  (u-u_\textrm{unc})^T M(u-u_\textrm{unc}).
\end{split}
\label{eq:u_quadratic}
\end{equation}
The feasible set $\mathcal U$ is known and described by linear inequality constraints. Therefore, this minimization is a simple convex quadratic program. 
Notice how both the unconstrained and the constrained solution do not depend on the unmeasured exogenous input $w$, as desired.

\subsection{Feedback Optimization for Volt/VAr Regulation}
\label{subsec:Volt/VAr}

In this section we specialize FO to the Volt/VAr regulation problem.
This problem is defined as follows: Determine the reactive power $q_h$ at every DER $h$ such that $q_\textrm{min}\leq q_h \leq q_\textrm{max}$ and that $v_\textrm{min} \leq v_h(q,w) \leq v_\textrm{max}$. Here, $v_h(q,w)$ is the steady state map of the nonlinear power flow equations that defines voltages $v_h$ as a function of both reactive powers $q_h$ and external influences $w$ (e.g., active and reactive demands, active generation). 
Mathematically speaking, we try to solve a feasibility problem:
\begin{equation*}
    q\in \mathcal{F} \quad \mathcal{F}:=\{q \;|\; q_\textrm{min}\leq q \leq q_\textrm{max},\; v_\textrm{min} \leq v(q,w) \leq v_\textrm{max}\} ,
\end{equation*}
where $q$ and $v$ are the vectors of reactive power set-points and voltage magnitudes that we obtain by stacking the individual $q_h$ and $v_h$ of the DERs, respectively.
We choose not to control active power with our algorithm. Due to the different cost of the two control actions one should first utilize reactive power and only afterwards use active power to control the voltage. Therefore, these two control actions can be applied individually and do not need a unified control approach. However, active power could easily be included in the controller without adding technical difficulties.
In order to apply the proposed methodology, we cast this feasibility problem into the optimization problem
\begin{equation}
    \begin{alignedat}{2}
        \min_q &\ \dfrac{1}{2}q^T M q\\
        \text{s.t.} &\  v_\text{min} \leq v_h(q,w) \leq v_\text{max} \quad &&\forall h\\
        &\ q_\text{min} \leq  q_h \leq q_\text{max} &&\forall h.
    \end{alignedat}
    \label{eq:reactive_optimization_problem}
\end{equation}
This is a special case of \eqref{eq:optimization_problem_qp}, where  $M$ can be used to weight the reactive power contribution of the different inverters $h$.

We introduce the dual variables $\lambda_\textrm{min}$ and $\lambda_\textrm{max}$ corresponding to the voltage (output) constraints.
We adapt \eqref{eq:lambda_update_linear} to this specific case (namely, 
$A = \left[\begin{smallmatrix} -I \\ I \end{smallmatrix}\right]$,
$b = \left[\begin{smallmatrix} v_\text{min} \\ -v_\text{max} \end{smallmatrix}\right]$) and we get
\begin{align}
    \lambda_\textrm{min}(t+1) &= [\lambda_\textrm{min}(t)+\alpha(v_\textrm{min}-v)]_{\geq 0}
        \label{eq:update_lambda_min}\\
    \lambda_\textrm{max}(t+1) &= [\lambda_\textrm{max}(t)+\alpha(v-v_\textrm{max})]_{\geq 0}
        \label{eq:update_lambda_max} .
\end{align}
As we can see, we are integrating the voltage violations, which can be measured, with a gain of $\alpha$.

As discussed before, in order to calculate \eqref{eq:u_unc}, we need a constant approximation of the sensitivity of the voltages with respect to the reactive power injection akin to power transfer distribution factors for active power generation on the transmission level.
Under no-load conditions and the assumption of negligible cable resistances we have the approximation
\begin{equation} \label{eq:linear_model}
    \frac{\partial v(q,w)}{\partial q} = X ,
\end{equation}
where $X$ is the reduced bus reactance matrix that can be derived from the grid topology and the data in Table \ref{tab:cable}. 
The approximation is accurate for lightly loaded systems, because the nonlinearity of the power flow equations is mild near this operating point \cite{Bolognani2015fast}.
In our application the system can be heavily loaded, but in Section~\ref{sec:experiments} we verify that the proposed FO is sufficiently robust against this model mismatch.

\begin{figure*}[ht!]
    \centering
    \includegraphics[width=0.9\textwidth]{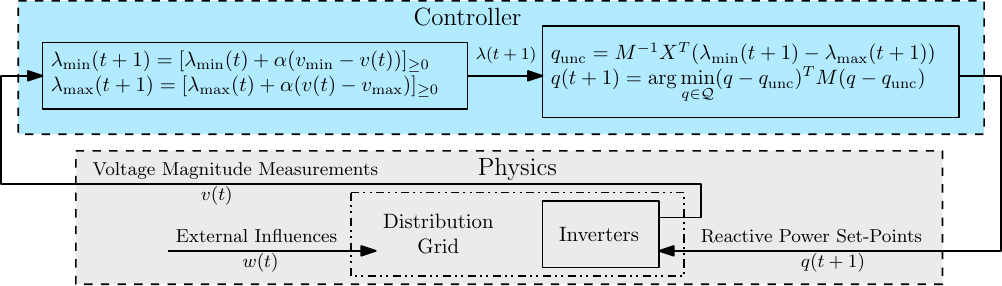}\\[2mm]
    \caption{Block diagram of the controller with \eqref{eq:update_lambda_min} and \eqref{eq:update_lambda_max} (left block) and \eqref{eq:q_unc} and \eqref{eq:q_quadratic} (right block). The controller gets the voltage magnitude measurements from the inverters and determines the reactive power set-points, which are send to the inverters. The parameter $\alpha$ is the controller gain and is the only tuning knob. Note, that the left block corresponds to the integral part of a PI-controller.}
    \label{fig:block_diagram}
\end{figure*}

The expression in \eqref{eq:u_unc} for the optimal unconstrained reactive power set-points $q_\textrm{unc}$ becomes
\begin{equation}
    q_\textrm{unc} = M^{-1}X^T (\lambda_\textrm{min}(t+1)-\lambda_\textrm{max}(t+1)),
        \label{eq:q_unc}
\end{equation}
while the solution of the constrained optimization problem \eqref{eq:u_quadratic} becomes
\begin{align}
    \begin{split}
        q(t+1) = \arg \min_{q \in \mathcal Q} \; (q-q_\textrm{unc})^T M(q-q_\textrm{unc}) ,
    \end{split}
        \label{eq:q_quadratic}
\end{align}
where  $\mathcal{Q}=\{q\;|\;q_\textrm{min}\leq q \leq q_\textrm{max}\}$.

In practice, these reactive power set-points $q(t+1)$ are to be communicated to the different DERs, which will adjust their reactive power accordingly and collect the measurement of the consequent steady state voltage magnitudes, which need to be communicated to the central control unit. Therefore, at every time step the measurement and set-point need to be communicated by and to every inverter, respectively.
The resulting centralized controller is represented in Figure~\ref{fig:block_diagram} and consists of equations \eqref{eq:update_lambda_min} and \eqref{eq:update_lambda_max} (left block in the figure) and \eqref{eq:q_unc} and \eqref{eq:q_quadratic} (right block in the figure).

We can see that the FO controller uses the same measurements as local controllers, but these measurement are processed by a central unit which coordinates the actions of the different DERs and steers the system to the optimal steady state.
In comparison to the OPF-based dispatch, no nonlinear model nor knowledge of the power consumption or generation (modelled as external influences $w$) is needed.

\section{Experimental Setup}
\label{sec:experimental_setup}

The experiment has been implemented in the SYSLAB distribution grid at DTU Ris\o, Denmark. A small yet realistic distribution feeder has been configured in order to observe an overvoltage condition caused by local generation. The same setup was used in \cite{Ortmann2020teamvar} to analyze a distributed FO controller for the Volt/VAr problem. Without proper reactive power control, the feeder's ability to host renewable energy injections is limited and generation has to be curtailed. This scenario was chosen because it constitutes a non-trivial voltage regulation problem which cannot be solved without a coordinated Volt/VAr control strategy, as will be demonstrated in Section \ref{subsubsec:droop}. Note, that the applicability of the proposed FO strategy is not limited to the chosen topology.

\begin{figure}
    \centering
    \includegraphics[width=0.9\columnwidth]{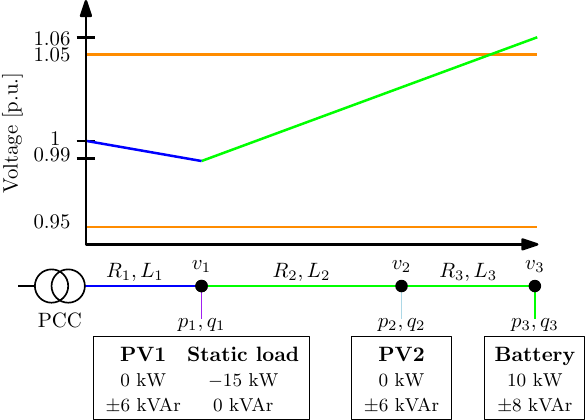}\\[2mm]
    \caption{Sketch of the voltage profile and the distribution feeder. The colors of the voltage profile match the colors of the sketched feeder.}
    \label{fig:voltage_profile}
\end{figure}

The setup consists of a vanadium battery, two photovoltaic systems (PV), a resistive load, and the distribution substation (PCC) connecting the distribution feeder to the remaining grid, see
Figure~\ref{fig:voltage_profile}.
The different nodes are connected via cables with non-negligible resistance (Table \ref{tab:cable}). The cable connecting the battery to the grid has a particularly large resistance.

The active power injection $p_3$ of the battery can represent a renewable source, which should not be curtailed. 
In our experiments we choose the active power of the battery to be $p_3=10$~kW. 
The high cable resistance and active power injection deteriorates the approximation of the sensitivity matrix in \eqref{eq:linear_model}. In Section \ref{subsubsec:FO_performance} we will show that the FO controller can cope with the model mismatch.
The static load is set to an active power consumption of 15 kW ($p_1=-15$~kW) which is larger than the local production, therefore requiring a positive active power flow from the substation.
PVs are fluctuating power sources. Therefore, to facilitate repeatability of the experiments and to allow for a comparison between different controllers, the PVs do not inject active power ($p_2=$~0kW).

The resulting voltage profile with no reactive power flows is represented in Figure~\ref{fig:voltage_profile}, where the overvoltage at the end of the feeder is apparent. 

Both the PVs and the battery can measure their voltage magnitudes, and their reactive power injections can be controlled.
The PV inverters have a reactive power range of $\pm$6~kVAr and the battery can be actuated with $\pm8$~kVAr.
The inverters at SYSLAB are oversized such that their full reactive power range is available independently of their concurrent active power injection.
The PVs and the battery can communicate with a central computational unit via a general-purpose Ethernet network, while the load is uncontrolled and unmeasured.

The voltage limits are defined to be 0.95~p.u. and 1.05~p.u.
We set these limits tighter than most grid codes in order to be able to observe persistent overvoltages without hardware protections being activated.

\begin{table}
\centering
\caption{Overview of the resistances and inductances in the grid.}
\label{tab:cable}
\begin{tabular}{cccccc}
\toprule
    $R_1 \;[\Omega]$ & $L_1 \;[\Omega]$ & $R_2 \;[\Omega]$ & $L_2 \;[\Omega]$ & $R_3 \;[\Omega]$ & $L_3 \;[\Omega]$ \\
    \midrule
     0.195 & 0.124 & 0.11 & 0.027 & 0.97 & 0.093\\
\bottomrule
\end{tabular}
\end{table}

\section{Controller Implementation}
\label{sec:controller_implementation}

The FO controller is implemented in Matlab at a central computation unit (Figure~\ref{fig:block_diagram}), where it is provided with the voltage magnitude measurements from the different inverters and computes the reactive power set-points. These are send to the inverters every 10 seconds, because the PV systems in the laboratory were not to be actuated more frequently, due to special hardware constraints. In general, the controller can run more frequently.

\subsection{Controller Tuning}
The controller has one tuning parameters which is the scalar control gain $\alpha$ in \eqref{eq:update_lambda_min} and \eqref{eq:update_lambda_max}. The higher its value, the faster a voltage constraint violation is integrated and the faster the DERs' reactive power set-points counteract the violation. However, as known from the optimization literature the stability of the gradient ascent we perform in \eqref{eq:generic_lambda_update} is lost if $\alpha$ is chosen too large (see \cite[Proposition 1.2.3]{Bertsekas1999}).

\subsection{Anti-Windup}

If the active power injections are too high (overvoltage) or too low (undervoltage) there do not exist feasible reactive power injections that lead to voltages which are inside the allowed voltage band. Therefore, the Volt/VAr problem is infeasible and at least one voltage violation is persistent. In this case the dual variable ($\lambda_\textrm{min}$ or $\lambda_\textrm{max}$) corresponding to the violated constraint keeps integrating,  yielding a windup of this variable. 
We implemented the following simple anti-windup solution in which the integration of the constraint violation is inhibited if all DERs are saturated:
\[
    \lambda_{h,\text{min}}(t+1) =
    \begin{cases}
    \lambda_{h,\text{min}}(t) 
    \hspace{22mm}\parbox[c][10mm][c]{27mm}{
        if
        $v_\text{min} - v_h(t) > 0$ \\
        and
        $q_k = q_{k,\text{max}}\ \forall k$
    } \\[-10pt]
    \lambda_{h,\text{min}}(t) + \alpha (v_\text{min} - v_h(t)) \hspace{8mm}
    \parbox[c][10mm][c]{15mm}{otherwise,}\\[-6pt]
    \end{cases}
\]
% \vspace{-0.4cm}
\[
    \lambda_{h,\text{max}}(t+1) =
    \begin{cases}
    \lambda_{h,\text{max}}(t) 
    \hspace{22mm}\parbox[c][10mm][c]{27mm}{
        if
        $v_h(t) - v_\text{max} > 0$ \\
        and
        $q_k = q_{k,\text{min}}\ \forall k$
    } \\[-10pt]
    \lambda_{h,\text{max}}(t) + \alpha (v_h(t) - v_\text{max}) \hspace{8mm}
    \parbox[c][10mm][c]{15mm}{otherwise.}\\[-6pt]
    \end{cases}
\]
Furthermore, an active power curtailment could be triggered once all DERs are saturated.
  
\section{Benchmark Controllers}

We implement a local droop controller and an OPF-based dispatch as two benchmark solutions to compare with the proposed FO strategy.
These approaches have almost opposite features: The droop controller only needs local voltage magnitude measurements, no communication, and no model of the grid; the OPF-based dispatch is centralized, requires communication of full state measurements (all power generation and demand), and relies on an accurate nonlinear grid model.

\subsection{Droop Control}

The droop controller that we implement complies with the recommendations by recent grid codes \cite{IEEE1547,VDE,ENTSOE}. 
Every DER measures the magnitude of the voltage at their point of connection and absorbs/injects reactive power following the piecewise linear control law

\vspace{3mm}
\noindent%
\begin{minipage}{0.6\columnwidth}
$\displaystyle
q_h \!=\!
    \begin{cases}
    q_{h,\text{max}}& v_h\! < \!v_1\\
    q_{h,\text{max}}\dfrac{v_2\!-\!v_h}{v_2\!-v_1} & v_1 \!\leq\! v_h \!\leq\! v_2\\
    0 & v_2\leq v_h\! \leq\! v_3\\
    q_{h,\text{min}}\dfrac{v_h\!-\!v_3}{v_4\!-\!v_3} & v_3 \! \leq \! v_h \!\leq\! v_4\\
    q_{h,\text{min}}& v_4\! < \!v_h.\\
    \end{cases}
$
\end{minipage}%
\begin{minipage}{0.4\columnwidth}
\includegraphics[width=\columnwidth]{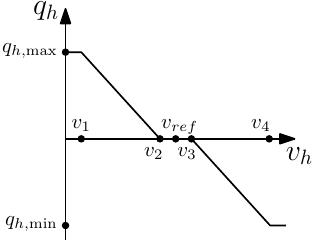}
\end{minipage}%
\vspace{3mm}
Based on the voltage band specifications of our experiment, we tune the droop curve to $v_1=0.95$~p.u., $v_2=0.99$~p.u., $v_3=1.01$~p.u. and $v_4=1.05$~p.u..

\subsection{OPF-based Dispatch}

We implement an OPF-based dispatch by communicating all reactive and active power consumption and generation to a centralized computation unit. There, we solve \eqref{eq:reactive_optimization_problem} using the OPF solver provided by Matpower \cite{Matpower}, which we provide with a nonlinear grid model that we obtain from the grid topology and the data from Table~\ref{tab:cable}. The reactive power set-points which are the solution of \eqref{eq:reactive_optimization_problem} are then given to the inverters. This approach guarantees optimality of the set-points under perfect model knowledge, but all power generation and consumption needs to be measured or estimated.
This information is available at SYSLAB with a significant level of accuracy.
In most distribution grids, the cable data and grid topology are not known exactly, nor are all reactive and active power consumption and generation measurements available.

\section{Experimental Results}
\label{sec:experiments}

In the following experiment, we analyze two crucial features: the tracking performance when solving a time-varying voltage regulation problem, and the robustness against model uncertainty.
We also contrast the proposed FO strategy with the local droop controller and the OPF-based dispatch.

\subsection{Tracking Performance}

We repeat the following 21-minute experiment for the three aforementioned strategies: droop control, OPF-based dispatch, and FO.
All power inverters are initialized with zero reactive power injection.\footnotemark 
\footnotetext{The plots show that the battery is injecting a small amount of reactive power at the beginning of the experiments. This is due to a measurement error. An inaccurate sensor is used for the internal reactive power controller of the battery, and a small tracking error is therefore present. The reported measurements in the figures are accurate.}

After three minutes the controllers are activated and start regulating the voltage. After 11 minutes the active power injection of the battery is reduced to 0~kW (effectively removing the cause of the overvoltage and the need for reactive power regulation). At minute 14 the active power injection is stepped up again to 10~kW for the remaining seven minutes of the experiment.

\subsubsection{Droop Control} \label{subsubsec:droop}
\begin{figure}
    \centering
	\begin{footnotesize}
    \input{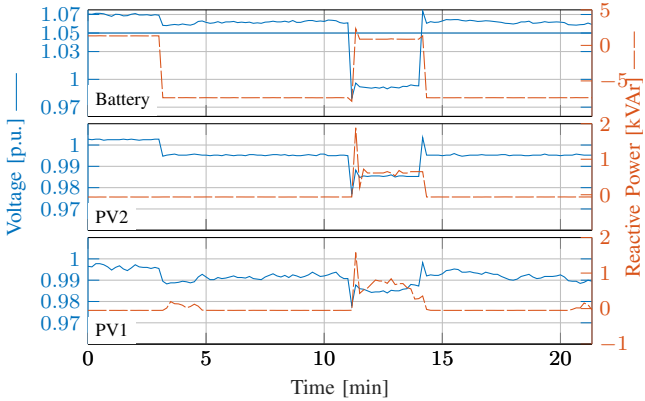}
    \caption{Performance of the Droop Control.}
    \label{fig:droop_normal}
    \end{footnotesize}
\end{figure}

The performance of the droop controller can be seen in Figure~\ref{fig:droop_normal}. Once the controller is activated the reactive power of the battery drops to its lower limit which reduces the overvoltage. However, the limited reactive power capability of the battery cannot  drive the voltage into the desired voltage range. The PV systems do not absorb reactive power to help reduce the overvoltage because they do not sense an overvoltage condition at their point of connection, and they will not lower their voltage below the nominal value of 1 p.u.
Using a lower nominal voltage is also not possible as it will increase the occurrence of undervoltage events.
This behavior is general for all local control strategies, and cannot be prevented without introducing some form of coordination between the inverters. Local control strategies are therefore inherently suboptimal; as established from a theoretical perspective in \cite{Bolognani2019}.

During minutes three to five, PV1 even injects reactive power to increase its voltage, because it has fallen under its deadband voltage of 0.99 p.u. This worsens the overvoltage at the battery, showing that droop control can even be detrimental.

\subsubsection{OPF-based Dispatch}

\begin{figure}
    \centering
	\begin{footnotesize}
    \input{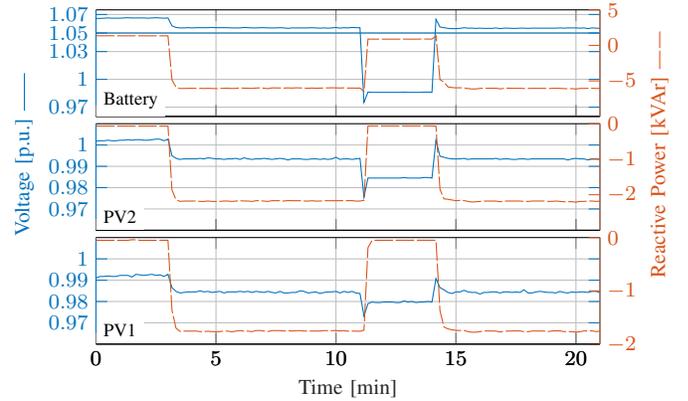}
    \caption{Performance of the OPF-based dispatch.}
    \label{fig:ORPF_normal}
    \end{footnotesize}
\end{figure}

An OPF-based strategy guarantees optimality under perfect model knowledge. This is a strong requirement which cannot be met in practice. Even in the SYSLAB distribution grid, where the setup, the cables and their parameters are accurately known, the OPF solution does not lead to feasible voltages (see the persistent voltage violation in Figure~\ref{fig:ORPF_normal}).
Standard techniques such as disturbance observers, model adaptation, and state estimation could be used to alleviate the effect of model uncertainty. Also, robust optimization techniques could be used to solve the OPF problem.
Nevertheless, an OPF-based dispatch requires a nonlinear grid model and knowledge of all active and reactive power consumption and production on the feeder. 

\subsubsection{Feedback Optimization}
\label{subsubsec:FO_performance}

\begin{figure}
    \centering
	\begin{footnotesize}
	\input{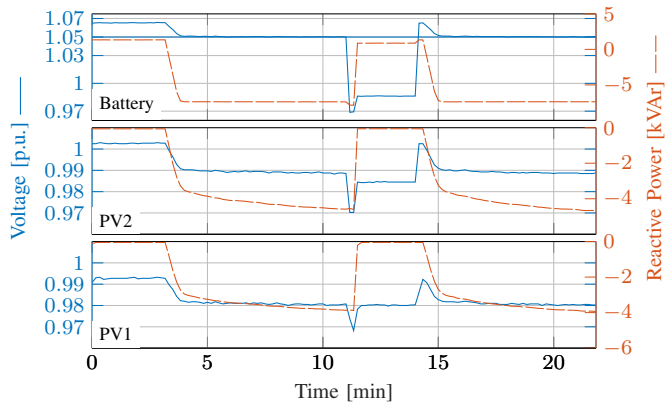}
    \caption{Performance of the FO controller.}
    \label{fig:network_normal}
    \end{footnotesize}
\end{figure}

The control gain $\alpha$ of the FO controller is chosen to be 100, and the matrix $X$ was calculated using the data from Table \ref{tab:cable}. The weighting matrix of the optimization problem $M$ is a diagonal matrix with the entries being the inverse of the reactive power limits ($q_\textrm{max}^{-1}$): 
\begin{equation*}
    X=
    \begin{bmatrix}
    0.10 & 0.09 & 0.09\\
    0.09 & 0.11 & 0.11\\
    0.09 & 0.11 & 0.16
    \end{bmatrix},
    \quad
    M=
    \begin{bmatrix}
    1/6 & 0 & 0\\
    0 &     1/6 & 0 \\
    0 & 0 &     1/8
    \end{bmatrix}.
\end{equation*}{}
The control performance can be seen in Figure~\ref{fig:network_normal}. When the controller is activated the central unit is provided with the voltages at the PV systems and the battery. The dual variable $\lambda_\textrm{max,3}$ that corresponds to the violation of the upper voltage limit of the battery starts integrating the violation.
This then leads to all inverters reducing their reactive power injections. As long as there is an overvoltage the dual variable keeps integrating, which leads to the inverter absorbing more reactive power which lowers the voltage. At steady state the voltage at the battery is at the upper voltage limit and the reactive power injections are at the optimal solution of \eqref{eq:reactive_optimization_problem}.

The temporal constraint violation before the system converges to the feasible voltage band can be made shorter by using a faster sampling time. Furthermore, the power system is equipped to withstand short overvoltages.

\subsection{Robustness to Model Mismatch}

\begin{figure}
    \centering
	\begin{footnotesize}
    \input{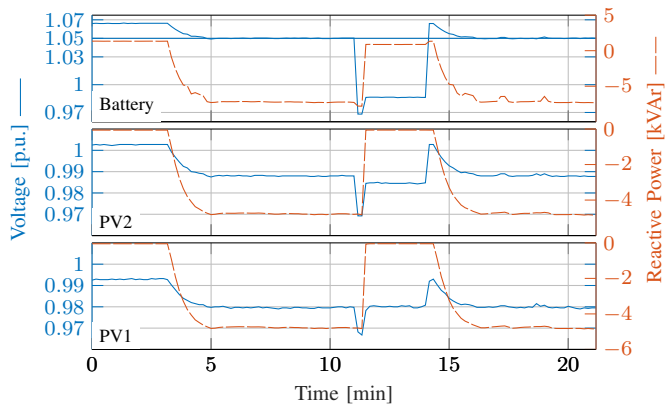}
    \caption{Performance of the FO controller with no model information.}
    \label{fig:network_normal_X_ones}
    \end{footnotesize}
\end{figure}

Due to its feedback nature, the proposed FO approach is expected to be robust to model mismatch. 
However, in spite of recent theoretical insights \cite{colombino2019towards},  the robustness of these strategies has not been analyzed experimentally before.
In order to validate this claim in an experiment, we assume uncertainty in the knowledge of the grid sensitivity matrix $X$.
We consider the crude approximation in which all entries of the $X$ matrix are believed to be $1$. This choice corresponds to assuming that all inverters are connected to the same point on the feeder.
No other model information is used, making the controller design essentially model-free.
The behavior of this FO controller with $\alpha=10$ can be seen in Figure~\ref{fig:network_normal_X_ones}. 
Notice, that the controller is still able to drive the voltages to the feasible voltage band. The DERs are utilized differently than in Figure~\ref{fig:network_normal} due to the different $X$ matrix. This leads to a different value of the cost function, which is within 12\% of the optimal value.

\section{Conclusion}
\label{sec:conclusion}

We have implemented three Volt/VAr control strategies on a real distribution feeder: local droop control, centralized OPF-based dispatch that guarantees optimal regulation under perfect model information, and a recently proposed FO scheme.
While the droop control fails to regulate voltages in a satisfactory manner (as predicted analytically), the OPF-based dispatch exhibits substantial fragility with respect to model uncertainty. 
In contrast, the FO strategy drives the system to the feasible voltage range while relying only on voltage measurements collected from the inverters (without measuring or estimating any power flows). Within our experimental setup, feedback optimization is extremely robust to model mismatch and its design and tuning is essentially model-free.

This leads us to conclude that feedback optimization is a promising approach for the real-time coordinated control of DERs in future  distribution grids.
We conjecture that these features of feedback optimization are not specific to this application and we plan to investigate them in the more general context of real-time control of power systems.

\section{Acknowledgment}

We thank Alexander Maria Prostejovsky and Kai Heussen for their support during the implementation of the experiments at DTU and for many useful inputs and fruitful discussions.

\bibliographystyle{IEEEtran}
\bibliography{IEEEabrv,pscc2020}

\end{document}